\documentclass[aps,tightenlines,nofootinbib,preprint,
superscriptaddress,groupedaddress,epsfig]{revtex4}

\usepackage{graphicx}

\begin{document}

\title{Is $D^{+}_{sJ}(2632)$ the first radial excitation of $D_{s}^{*}(2112)$? }

\vspace*{2cm}

\author{Chao-Hsi Chang}
\email{zhangzx@itp.ac.cn} \affiliation{CCAST (World Laboratory),
P.O.Box 8730, Beijing 100080, China\footnote{Not mailing
address.}}\affiliation{Institute of Theoretical Physics, Chinese
Academy of Sciences, P.O.Box 2735, Beijing 100080, China.}

\author{C. S. Kim}
\email{cskim@yonsei.ac.kr} \affiliation{Department of Physics,
Yonsei University, Seoul 120-749, Korea}
\author{Guo-Li Wang}
\email{glwang@cskim.yonsei.ac.kr;
glwang@itp.ac.cn}\affiliation{Institute of Theoretical Physics,
Chinese Academy of Sciences, P.O.Box 2735, Beijing 100080, China.}
 \affiliation{Department of
Physics, Yonsei University, Seoul 120-749, Korea}
\affiliation{Department of Physics, Harbin Institute of
Technology, Harbin 150006, China}

 \baselineskip=20pt


\begin{abstract}

\noindent We present a quantitative analysis of the
$D^{+}_{sJ}(2632)$ observed by SELEX mainly focusing on the
assumption that $D^{+}_{sJ}(2632)$ is the first radial excitation
of the $1^{-}$ ground state $D^{*}_{s}(2112)$. By solving the
instantaneous Bethe-Salpeter equation, we obtain the mass $2658\pm
15$ MeV for the first excited state, which is about $26$ MeV
heavier than the experimental value $2632\pm 1.7$ MeV. By means of
PCAC and low-energy theorem we calculate the transition matrix
elements and obtain the decay widths:
$\Gamma(D^{+}_{sJ}\rightarrow D^{+}_s\eta)=4.07\pm 0.34$ MeV,
$\Gamma(D^{+}_{sJ}\rightarrow D^{0}K^{+}) \simeq
\Gamma(\Gamma(D^{+}_{sJ}\rightarrow D^{+}K^{0})=8.9\pm 1.2$ MeV,
and the ratio $\Gamma(D^{+}_{sJ}\rightarrow
D^{0}K^{+})/\Gamma(D^{+}_{sJ}\rightarrow D^{+}_{s}\eta)=2.2\pm
0.2$ as well. This ratio is quite different from the SELEX data
$0.14\pm 0.06$. The summed decay width of those three channels is
approximately $21.7$ MeV, already larger than the observed bound
for the full width ($\leq 17$ MeV). Furthermore, assuming
$D_{sJ}^+(2632)$ is  $1^{-}$ state, we also explore the
possibility of $S-D$ wave mixing to explain the SELEX observation.
Based on our analysis, we suspect that it is too early to conclude
that $D^{+}_{sJ}(2632)$ is the first radial excitation of the
$1^{-}$ ground state $D^{*}_{s}(2112)$. More precise measurements
of the relative ratios and the total decay width are urgently
required especially for $S-D$ wave mixing.

\end{abstract}


\maketitle

\section{Introduction}

Last year the SELEX Collaboration reported the first observation
of a narrow charmed meson $D^{+}_{sJ}(2632)$ with the mass of
$2632.5\pm 1.7$ MeV and decay width $\Gamma_{tot} \leq 17$ MeV at
$90\%$ confidence level \cite{selex}. This state has been seen in
two decay modes, $D^{+}_{s}\eta$ and $D^{0}K^{+}$, with a relative
branching ratio $\Gamma(D^{+}_{sJ}\rightarrow
D^{0}K^{+})/\Gamma(D^{+}_{sJ}\rightarrow D^{+}_{s}\eta)=0.14\pm
0.06$. This relative branching ratio is rather unusual
because the phase space of channel $D^{0}K^{+}$ is $1.5$ times
larger than that of the channel $D^{+}_{s}\eta$ and according to
the SU(3) flavor symmetry the ratio $\Gamma(D^{+}_{sJ}\rightarrow
D^{0}K^{+})/\Gamma(D^{+}_{sJ}\rightarrow D^{+}_{s}\eta)\geq 2.3$
\cite{chunliu1} if the quark content of $D^{+}_{sJ}(2632)$ is
$c\bar s$, and the reference \cite{feclose} gave an estimate of
$\Gamma(D^{+}_{sJ}\rightarrow
D^{0}K^{+})/\Gamma(D^{+}_{sJ}\rightarrow D^{+}_{s}\eta)=7.0$.
Since the $D^{+}_{sJ}(2632)$ is above the threshold for
$D^{0}K^{+}$ and $D^{+}_{s}\eta$, and its decays to these final
states are Okubo-Zweig-Iizuka (OZI) allowed processes, it is also
strange that it has a narrow width ($\leq 17$ MeV).

Why this state is so narrow and dominated by the $D^{+}_{s}\eta$
decay mode has inspired a lot of theoretical interest
\cite{chunliu1,feclose,tjon,ktchao,chunliu2,xueqianli,
ailinzhang,beveren,maiani,nowak,godfrey2,nicolescu,gupta,close2},
with many possible explanations, $e.g.$ the first radial
excitation of $D^{\star}_{s}(2112)$, a tetraquark, a hybrid, a
two-meson molecule, a diquark-antidiquark bound state, $etc.$
Among them the most plausible candidate is the first radial
excitation of $D^{\star}_{s}(2112)$, the $2 {^3}S_{1}$ state,
since it is the likely conventional meson with component of $c\bar
s$ which has not been found. And even though the mass of $2632$
MeV is about $100$ MeV below the traditional potential model
prediction (for example, see \cite{poten}), the similar situation
has been found in the new narrow charm-strange $0^{+}$ meson
$D_{sJ}(2317)$, favored as the conventional $c \bar s$ state.
Furthermore, Simonov and Tjon \cite{tjon} showed a possible mechanism to shift
down the meson mass by the coupled-channel analysis method. Chao
\cite{ktchao} gave a reasonable argument that the node structure in
the $2 {^3}S_{1}$ state may explain the small decay width and
unusual decay modes.

In this letter, following the references \cite{feclose,ktchao},
we assume $D^{+}_{sJ}(2632)$ as the first
radial excitation of vector meson $D_{s}^{*}(2112)$.
Then we solve the instantaneous
Bethe-Salpeter equation \cite{BS}, and give our prediction of mass
for $2{^3}S_{1}$ state. At the same time, we obtain the wave
functions of the relevant mesons. By using the reduction formula,
PCAC relation and low energy theorem, we write the transition
$S$-matrix as a formula involving the light meson decay constant
and the corresponding axial current transition matrix element
between two heavy mesons, which in turn can be written as an
overlapping integral of the relevant wave functions. We give
a detailed consideration on the
node structure in $2^3 S_1$ wave function.
We further try to make a rough estimate on the full decay width and the ratio
$\Gamma(D^{+}_{sJ}\rightarrow D^{0}K^{+})/\Gamma(D^{+}_{sJ}\rightarrow D^{+}_{s}\eta)$
by assuming $D^{+}_{sJ}(2632)$ as the $D$-wave $1^{-}$ vector state as well as considering
the possible $S-D$ wave mixing.

\section{Prediction of Mass}

In our previous papers \cite{cskimwang,changwang2}, we solved
exactly the instantaneous Bethe-Salpeter equations (or the
Salpeter equations \cite{salp}) to describe the behavior of
$0^{-}$ and $1^{-}$ states. In Refs.  \cite{cskimwang}, we have
given the mass spectra and wave functions of pseudoscalar $D_s$,
$D^{0}$ and $D^{+}$, and the mass spectra of all these heavy
$0^{-}$ states, which are consistent with experimental data quite
well. For the instantaneous Bethe-Salpeter equation of $1^{-}$
states \cite{changwang2}, we have used the same parameters as used
in \cite{cskimwang}, only changing the parameter $V_0$ to fit data
because $V_0$ is the only parameter which can be different for
$0^{-}$ and $1^{-}$ states when their quark contents are the same. In
this letter we first solve the full Salpeter $1^{-}$ equation
through fitting the ground state mass at $2112$ MeV
($D^{*}_{s}(2112)$) to fix the parameter $V_0$. After fixing the
parameter $V_0$, we then give the predictions of the first radial
excitation state. Since we are only interested in the leading
order estimate, which lets us  see the problem clearly,
we assume in our calculation that the first radial excitation state
$D_{sJ}^{+}$ is a pure $S$ wave state, excluding the possible
mixing between $S$ wave and $D$ wave states. Later in this letter we will
also consider the $D$ wave contribution. Our prediction for the mass of
the first radial excitation state is $2658\pm 15$ MeV, which is
about $26$ MeV heavier than the experimental data $2632\pm 1.7$
MeV.

\section{Transition matrix element}

In this section we give a detailed description for the calculation of the
transition matrix element. By using the reduction formula, the
transition $S$-matrix of decay $D^{+}_{sJ}\rightarrow D^{0}K^{+}$
(similar to the channel $D^{+}_{sJ}\rightarrow D^{+}K^{0}$) can be
written as:
\begin{equation}
\langle D(P_{f1})K(P_{f2})|D^{+}_{sJ}(P)\rangle= \int d^{4}x
e^{iP_{f2}\cdot x}(M^2_{_K}-P^{2}_{f2})\langle
D(P_{f1})|\Phi_{K}(x)|D^{+}_{sJ}(P)\rangle,
\end{equation}
where $P$, $P_{f1}$ and $P_{f2}$ are the total momenta of initial
state $2^3S_1$, final state heavy meson and final state light
meson, respectively (see Figure 1); $M_K$ is the mass of the final
light meson. The PCAC relates the meson field $\Phi_{K}(x)$
with a current
$\Phi_{K}(x)=\frac{1}{M^{2}_{K}f_{_K}}\partial^{\mu}(\bar{q}\gamma_{\mu}\gamma_{5}s)$,
where $f_{_K}$ is the decay constant of light meson; $q$ is $u$
for $K^{+}$, $d$ for $K^{0}$, respectively. Then the $S$-matrix
becomes
$$\langle D(P_{f1})K(P_{f2})|D^{+}_{sJ}(P)\rangle
=\frac{(M^{2}_{K}-P^{2}_{f2})}{M^{2}_{K}f_{_K}}\int d^{4}x
e^{iP_{f2}\cdot x}\langle
D(P_{f1})|\partial^{\mu}(\bar{q}\gamma_{\mu}\gamma_{5}s
)|D^{+}_{sJ}(P)\rangle$$
\begin{equation}
=\frac{-iP^{\mu}_{f2} (M^{2}_{K}-P^{2}_{f2})}{M^{2}_{K}f_{_K}}\int
d^{4}x e^{iP_{f2}\cdot x}\langle
D(P_{f1})|\bar{q}\gamma_{\mu}\gamma_{5}s |D^{+}_{sJ}(P)\rangle,
\label{pcac1}
\end{equation}
and with the low energy theorem this equation can be written
approximately as:
\begin{equation}
\approx \frac{-iP^{\mu}_{f2} }{f_{_K}}\int d^{4}x e^{iP_{f2}\cdot
x}\langle D(P_{f1})|\bar{q}\gamma_{\mu}\gamma_{5}s
|D^{+}_{sJ}(P_i)\rangle.
\end{equation}
And, finally, we have
\begin{eqnarray}
\langle D(P_{f1})K(P_{f2})|D^{+}_{sJ}(P_i)\rangle
=(2\pi)^{4}\delta^{4}(P_i-P_{f1}-P_{f2})\frac{-iP_{f2}^{\mu}}{f_{_K}}
\langle
D(P_{f1})|\bar{q}\gamma_{\mu}\gamma_{5}s|D^{+}_{sJ}(P_i)\rangle.
\label{eq1}
\end{eqnarray}

\begin{figure}
\begin{picture}(100,170)(300,435)
\put(0,0){\includegraphics{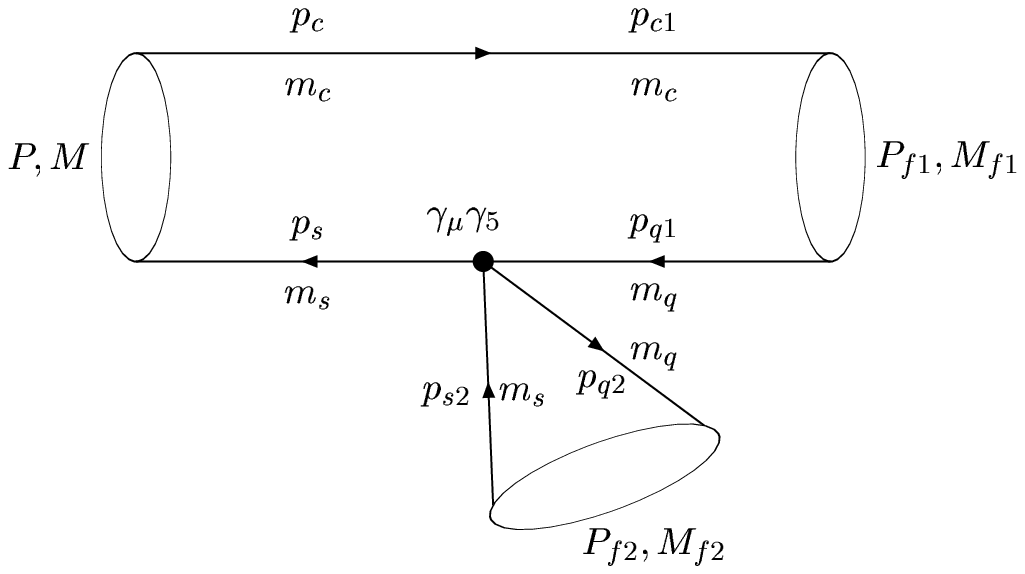}}
\end{picture}
\caption{Diagram for the transition of $D^{+}_{sJ}(2632)$ decay. }
\end{figure}

For the decay $D^{+}_{sJ}\rightarrow D^{+}_{s}{\eta}$, the
equation is similar, but there is $\eta - \eta'$ mixing.
Including the $\eta - \eta'$ mixing, the PCAC relation reads
as
$$\Phi_{\eta}(x)=\cos\theta\Phi_{\eta_8}(x)+\sin\theta\Phi_{\eta_0}(x)$$
$$=\frac{\cos\theta}{M^{2}_{\eta_8}f_{\eta_8}}
\partial^{\mu}(\frac{\bar{u}\gamma_{\mu}\gamma_{5}u+
\bar{d}\gamma_{\mu}\gamma_{5}d
-2\bar{s}\gamma_{\mu}\gamma_{5}s}{\sqrt 6}) +
\frac{\sin\theta}{M^{2}_{\eta_0}f_{\eta_0}}
\partial^{\mu}(\frac{\bar{u}\gamma_{\mu}\gamma_{5}u+
\bar{d}\gamma_{\mu}\gamma_{5}d
+\bar{s}\gamma_{\mu}\gamma_{5}s}{\sqrt 3})$$
\begin{equation}
=\partial^{\mu}(\bar{s}\gamma_{\mu}\gamma_{5}s)\left[\frac{-2\cos\theta}{{\sqrt
6}M^{2}_{\eta_8}f_{\eta_8}}+\frac{\sin\theta}{{\sqrt
3}M^{2}_{\eta_0}f_{\eta_0}}\right],
\end{equation}
where $f_{\eta_8}$, $f_{\eta_0}$ are the decay constants of octet
$\eta_8$ and singlet $\eta_0$, respectively. Then the transition matrix
element becomes
$$
\langle D_s(P_{f1})\eta(P_{f2})|D^{+}_{sJ}(P_i)\rangle
=(2\pi)^{4}\delta^{4}(P_i-P_{f1}-P_{f2})({-iP_{f2}^{\mu}})$$
\begin{eqnarray}\times\left[
\frac{-2M^{2}_{\eta}\cos\theta}{{\sqrt
6}M^{2}_{\eta_8}f_{\eta_8}}+\frac{M^{2}_{\eta}\sin\theta}{{\sqrt
3}M^{2}_{\eta_0}f_{\eta_0}}\right] \langle
D_s(P_{f1})|\bar{s}\gamma_{\mu}\gamma_{5}s|D^{+}_{sJ}(P_i)\rangle.
\label{pcac2}
\end{eqnarray}
Note that the decays $D^{+}_{sJ}(2632)\to D^+_s$ are
OZI-rule-allowed, so we may ignore anomaly of the PCAC safely in
Eqs. (\ref{pcac1},\ref{pcac2}).

According to the Mandelstam formalism  \cite{mandelstam}, at the
leading order, the matrix element $\langle
D(P_{f1})|\bar{q}\gamma_{\mu}\gamma_{5}s|D^{+}_{sJ}(P_i)\rangle$
can be written as \cite{changwang}:
\begin{eqnarray}
\langle D(P_{f1})|\bar{q}\gamma_{\mu}\gamma_{5}s|D^{+}_{sJ}(P)\rangle=
\int\frac{d{\vec q}}{(2\pi)^3}Tr\left[
\bar{\varphi}^{++}_{_{P_{f1}}}({ \vec q}-\frac{m_c}{m_c +m_q}{\vec
r })\frac{\not\!P}{M} {\varphi}^{++}_{_{P_{i}}}({\vec
q})\gamma_{\mu}\gamma_{5} \right],\label{eq2}
\end{eqnarray}
where $m_c$ and $m_q$ is the mass of $c$ and $q$ ($q=u,d,s$)
quark; $\vec r$ is the recoil three dimensional momentum of the final
state $D$ meson. $p_c$, $p_s$ are the momenta of $c$  and $s$
quark in the initial meson, respectively. And the relative
momentum $\vec q$ is defined as $\vec{q}\equiv
\vec{p}_c-\frac{m_c}{m_c+m_s}\vec{P}\equiv
\frac{m_s}{m_c+m_s}\vec{P}-\vec{p}_s$, in the center of mass
system of the initial meson, and it becomes $\vec{q}\equiv
\vec{p}_c\equiv -\vec{p}_s$; ${\varphi}^{++}_{_{P_{i}}}$, and
${\varphi}^{++}_{_{P_{f1}}}$ are called  the positive energy wave
function of the initial and final heavy mesons, and
$\bar{\varphi}^{++}_{_{P_{f1}}}=-\gamma_{0}({\varphi}^{++}_{_{P_{f1}}})^{+}\gamma_{0}$.
The positive energy wave function of $D^{0}$ can be found in paper
\cite{cskimwang}:
$$
\varphi^{++}_{_{P_{f1}}}(\stackrel{\rightarrow}{q})=
\frac{M_{f1}}{2}\left(\varphi_1(\stackrel{\rightarrow}{q})
+\varphi_2(\stackrel{\rightarrow}{q})\frac{m_{c}+m_{q}}{\omega_{c}+\omega_{q}}\right)
$$\begin{equation}\times\left[
\frac{\omega_{c}+\omega_{q}}{m_{c}+m_{q}}+{\gamma_{_0}}-\frac{{\not\!
\vec q}(m_{c}-m_{q})}
{m_{q}\omega_{c}+m_{c}\omega_{q}}+\frac{{\not\! \vec
q}\gamma_{_0}(\omega_{c}
+\omega_{q})}{(m_{q}\omega_{c}+m_{c}\omega_{q})}\right]\gamma_{_5}\;,
\end{equation}
where $\omega_{c}=\sqrt{m_{c}^{2}+{\vec q}^{2}}$ and
$\omega_{q}=\sqrt{m_{q}^{2}+{\vec q}^{2}}$;
$\varphi_1(\vec{q})$, $\varphi_2(\vec{q})$ are the radial part wave
functions, and their numerical values can be obtained by solving the
full Salpter equation of $0^{-}$ state.

Since the wave function ${\varphi}^{++}_{_{P_{i}}}({\vec q})$
directly relates to the first radially excited state, there is
only one node in it. We note that the node structure can possibly
play an important role to explain both the narrow decay width and
the dominance of $D^{+}_{s}\eta$ decay mode, to a certain degree.
The value of the radial wave function of $2^3S_1$ state becomes
negative as a function of relative momentum $\vec q$ when it
crosses the node (denoted as $q_0$), that is when $|\vec q|>q_0$.
Therefore, this negative part of the wave function tends to cancel
the contribution to the decay width from the positive part of the
same wave function, which can be the reason to have a small decay
width. In the overlapping integral of the wave functions Eq.
(\ref{eq2}) there is the shift of momentum, ${\vec
q}-\frac{m_c}{m_c +m_q}\vec{r}$, and this will even strengthen the
negative contribution because the shift will move the peak of the
radial wave function of the final $D$ meson close to the node. We
note that the value of the overlapping integral will be
significantly affected by the momentum shift as the larger
momentum shift will cause the larger negative contribution, and in
turn the smaller overlapping integral (smaller decay width). Since
the recoil momentum $|\vec r|$ of $D$ meson is larger than that of
$D_s$, about $1.43$ times, and the momentum shift in case of $DK$
is $1.56$ times larger than that of the $D_{s}\eta$ channel,
larger negative contribution results in $DK$ channel than in the
$D_{s}\eta$ channel, which can be the very reason for the small
ratio of the branching ratios $DK/D_{s}\eta$.

As already pointed out in Ref. \cite{feclose}, there may be the $S-D$
mixing in $1^{-}$ state \cite{feclose,changwang2},
so we will try to estimate the possible mixing contributions here too.
We first assume that this $1^{-}$ state is a pure $S$ wave state
$2 ^3S_1$, and later we will consider the case of $D$ wave state.
The wave function of $1^{-}$ can be written as \cite{changwang2}:
$$
\varphi_{_{P_{i}}}(\vec {q})=M\left\{
(\psi_{1}+\psi_{2}\gamma_0){\not\!\epsilon}+\gamma_{0}\left[
\gamma_{0}{\vec q}\cdot{\epsilon}\psi_{1}+ ({\not\!\epsilon}
{\not\! \vec q}-{\vec
q}\cdot{\epsilon})\psi_{2}\right]\frac{(\omega_{c}\omega_{q}+{\vec
q}^2-m_{c}m_{q})} {(m_c+m_q){\vec q}^2} \right.$$
\begin{eqnarray}-\left.\left[
\gamma_{0}{\vec q}\cdot{\epsilon}\psi_{2}+ ({\not\!\epsilon}
{\not\! \vec q}-{\vec q}\cdot{\epsilon})\psi_{1}\right]
\frac{(m_{q}\omega_{c}-m_{c}\omega_{q})}{(\omega_{c}
+\omega_{q}){\vec{q}}^2} \right\},
\end{eqnarray}
and the positive energy wave function is defined as:
\begin{eqnarray}
\varphi^{++}_{_{P_{i}}}(\vec
{q})=\Lambda^{+}_{c}\gamma_{0}\varphi_{_{P_{i}}}(\vec
{q})\gamma_{0}\Lambda^{+}_{q}
=\frac{1}{2\omega_{c}}(\omega_{c}\gamma_{0}+m_c+{\not\! \vec
q})\gamma_{0} \varphi_{_{P_{i}}}(\vec
{q})\gamma_{0}\frac{1}{2\omega_{q}}(\omega_{q}\gamma_{0}-m_q-{\not\!
\vec q}),
\end{eqnarray}
where $\epsilon$ is the polarization vector of a $^3S_1$ state;
$\Lambda^{+}_{c}$, $\Lambda^{+}_{q}$ are the energy projection
operators for quark and antiquark; the numerical value of the
radial part wave function $\psi_1(\vec{q})$, $\psi_2(\vec{q})$
will be obtained by solving the exact Salpter equation of $1^{-}$
state.

\section{Numerical Results and Discussions}

For our numerical calculations we use the following
parameters: $m_c=1755.3$ MeV, $m_s=487$ MeV, $m_d=311$ MeV,
$m_u=305$ MeV \cite{cskimwang}, $f_{\pi}=130.7$ MeV
$f_{\eta_8}=1.26f_{\pi}$ MeV, $f_{\eta_0}=1.07f_{\pi}$ MeV
\cite{feldmann} and $f_{_K}=159.8$ MeV  \cite{pdg2004}. The mass of
octet $\eta_8$ can be estimated by the Gell-Mann-Okubo relation
$3m^2_{\eta_8}=4m^2_{_K}-m^2_{\pi}$, the mass of singlet $\eta_0$
can be obtained by
$m^2_{\eta_0}=m^2_{\eta}+m^2_{\eta'}-m^2_{\eta_8}$, and the mixing
angle can be estimated by the relation $m^2_{\eta_0}=\sin^2\theta
m^2_{\eta}+\cos^2\theta m^2_{\eta'}$. [So our input values are
$m_{\eta_8}=564.3$ MeV, $m_{\eta_0}=948.1$ MeV,
$\theta=-9.95^{\circ}$.]

We estimate the mass of the first radial
excitation of $D^{*}_{s}(2112)$ by solving the full Salpeter
equation of $1^{-}$ state, and our prediction is $2658\pm 15$ MeV,
which is $26$ MeV heavier than the central value
$2632\pm 1.7$ MeV measured by SELEX.
The theoretical uncertainties are estimated by varying all the
input parameters simultaneously within $\pm 5\%$, and we choose the
largest possible error. We obtained the wave
functions of the corresponding states, and we show radial part
wave functions $\varphi_1$, $\varphi_2$ of $D_s$ in Figure 2;
$\varphi_1$, $\varphi_2$ of $D^{0}$ in Figure 3; $\psi_1$,
$\psi_2$ of $2 ^3S_1$ in Figure 4. We show the overlapping part of
$2 ^3S_1$ wave function $\psi_{1}(|\vec q|)$ and $D_s$ wave
function $\varphi_{1}(|\vec q|-\frac{m_c}{m_c +m_s}{|\vec
r_{_{D_{s}}}|})$, when it has the biggest momentum shift
$\frac{m_c}{m_c +m_s}{|\vec r_{_{D_{s}}}|}$ in Figure 5. We also
show $\psi_{1}(|\vec q|)$ and $D^{0}$ wave function
$\varphi_{1}(|\vec q|-\frac{m_c}{m_c +m_u}{|\vec r_{_{D^{0}}}|})$
in Figure 6.  One can see that in $2 ^3S_1$ wave function the node
localizes at $q_0=0.54$ MeV, and the biggest momentum shifts
are $\frac{m_c}{m_c +m_u}|\vec{r_{_{D^{0}}}}|=0.447$ MeV  for $D^{0}$, and
$\frac{m_c}{m_c +m_s}|\vec{r_{_{D_{s}}}}|=0.283$ MeV for $D^{+}_{s}$.

The decay width for the two body final state can be written as:
\begin{equation}
\Gamma=\frac{1}{8\pi}\frac{|\vec r|}{M^2}|T|^2.
\end{equation}
Here the matrix element $T$ for $D^{+}_{sJ}\rightarrow D^{0}K^{+}$
and $D^{+}_{sJ}\rightarrow D^{+}K^{0}$ is
\begin{equation}
T=\frac{P_{f2}^{\mu}}{f_{_{P_{f2}}}}\langle
D(P_{f1})|\bar{q}\gamma_{\mu}\gamma_{5}s|D^{+}_{sJ}(P_i)\rangle ,
\end{equation}
and for $D^{+}_{sJ}\rightarrow D^{+}_s\eta$ it is
\begin{equation}T={P_{f2}^{\mu}}\left[
\frac{-2M^{2}_{\eta}\cos\theta}{{\sqrt
6}M^{2}_{\eta_8}f_{\eta_8}}+\frac{M^{2}_{\eta}\sin\theta}{{\sqrt
3}M^{2}_{\eta_0}f_{\eta_0}}\right] \langle
D(P_{f1})|\bar{q}\gamma_{\mu}\gamma_{5}s|D^{+}_{sJ}(P_i)\rangle .
\end{equation}
We obtain the following decay widths:
\begin{equation}
\Gamma(D^{+}_{sJ}\rightarrow D^{+}_s\eta)=4.07 \pm 0.34 \text{
MeV},
\end{equation}
\begin{equation}
\Gamma(D^{+}_{sJ}\rightarrow D^{0}K^{+})=8.87 \pm 1.23 \text{
MeV},
\end{equation}
\begin{equation}
\Gamma(D^{+}_{sJ}\rightarrow D^{+}K^{0})=8.74 \pm 1.22 \text{
MeV}.
\end{equation}
And the corresponding ratios are
\begin{equation}
\Gamma(D^{+}_{sJ}\rightarrow
D^{0}K^{+})/\Gamma(D^{+}_{sJ}\rightarrow D^{+}_{s}\eta)\simeq
\Gamma(D^{+}_{sJ}\rightarrow
D^{+}K^{0})/\Gamma(D^{+}_{sJ}\rightarrow D^{+}_{s}\eta) \simeq
2.2\pm 0.2.
\end{equation}
As can be seen, we cannot find the dominance of $D_s\eta$, and the ratio is much
different from the experimental ratio $0.14\pm 0.06$. The sum
of those three decay widths is $21.7$ MeV, already larger than the observed
bound for full width ($\leq 17$ MeV) given by SELEX experiment.
Therefore, our assumption of
the pure $S$ wave vector state cannot fit the
experimental data.

\begin{figure}
\begin{picture}(400,170)(0,-5)
\put(0,0){\includegraphics{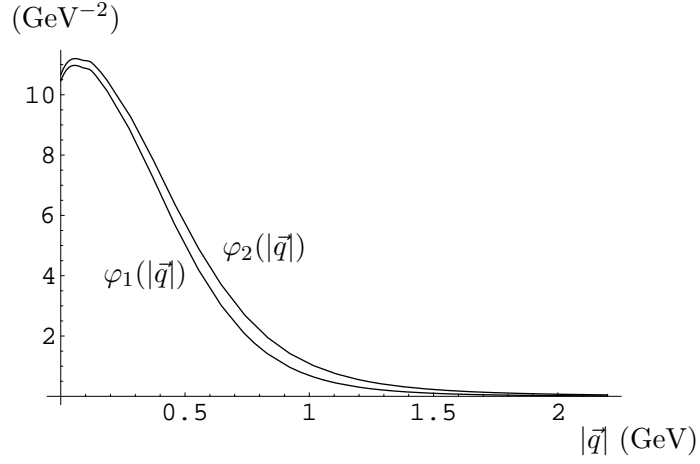}}
\put(305,-3){\makebox(2,2)[l]{$|\vec q| \text{ (GeV)}$}}
\put(90,157){\makebox(2,2)[l]{(GeV$^{-2}$)}}
\put(170,70){\makebox(2,2)[l]{$\varphi_{2}(|\vec q |)$}}
\put(125,60){\makebox(2,2)[l]{$\varphi_{1}(|\vec q |)$}}
\end{picture}
\caption{Radial wave function $\varphi_{1}(|\vec q |)$ and
$\varphi_{2}(|\vec q |)$ of $D^{+}_{s}$. }
\end{figure}

\begin{figure}
\begin{picture}(400,170)(0,-5)
\put(0,0){\includegraphics{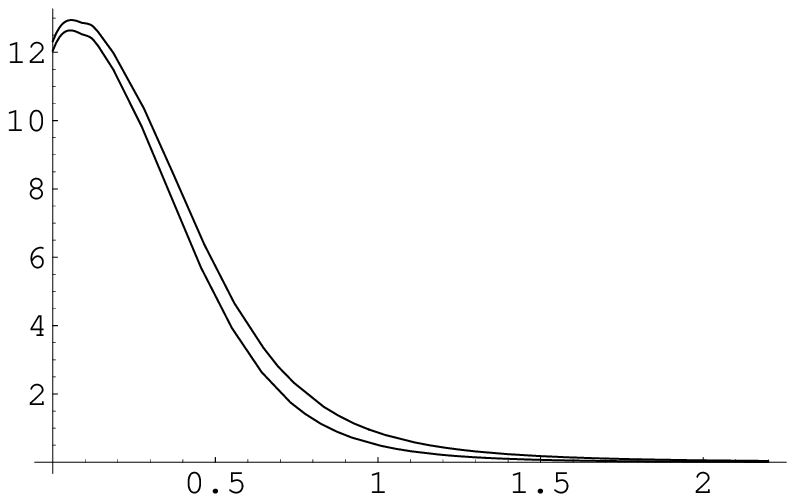}}
\put(305,-3){\makebox(2,2)[l]{$|\vec q| \text{ (GeV)}$}}
\put(90,157){\makebox(2,2)[l]{(GeV$^{-2}$)}}
\put(165,70){\makebox(2,2)[l]{$\varphi_{2}(|\vec q |)$}}
\put(120,60){\makebox(2,2)[l]{$\varphi_{1}(|\vec q |)$}}
\end{picture}
\caption{Radial wave function $\varphi_{1}(|\vec q |)$ and
$\varphi_{2}(|\vec q |)$ of $D^{0}$. }
\end{figure}

\begin{figure}
\begin{picture}(400,170)(0,-5)
\put(0,0){\includegraphics{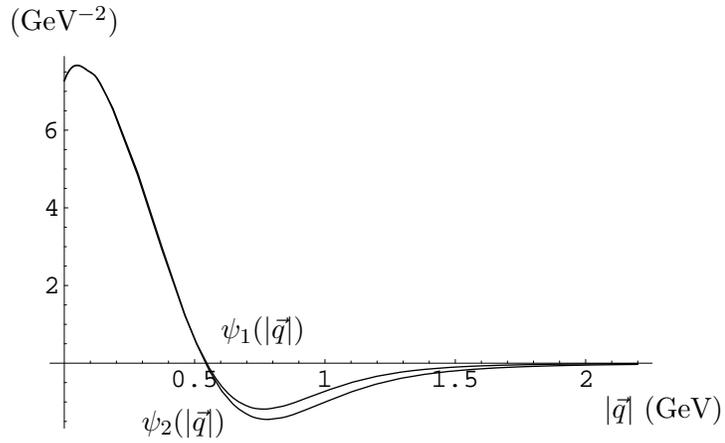}}
\put(305,10){\makebox(2,2)[l]{$|\vec q| \text{ (GeV)}$}}
\put(80,157){\makebox(2,2)[l]{(GeV$^{-2}$)}}
\put(160,40){\makebox(2,2)[l]{$\psi_{1}(|\vec q |)$}}
\put(130,5){\makebox(2,2)[l]{$\psi_{2}(|\vec q |)$}}
\end{picture}
\caption{Radial wave function $\psi_{1}(|\vec q |)$ and
$\psi_{2}(|\vec q |)$ of $2 ^3S_1$.}
\end{figure}

Before considering the $S-D$ mixing in $1^-$ vector state, let us first
assume $D^{+}_{sJ}(2632)$ to be a pure $D$ wave state. In this case, we
do not solve the full Salpeter equation, but instead, we give the leading
order estimate for the pure $D$ wave explanation. The wave
function of a pure $D$ wave state ($S-L$ coupling) can be
simplified as:
\begin{eqnarray}
\varphi_{_{P_{i}}}(\vec
{q})=\sum_{s_z,l_z}M(1+\gamma_0){\not\!\epsilon}(s_z)\psi_{nll_z}(\vec
q)<1s_{z};ll_{z}\mid 1j_{z}>\,,
\end{eqnarray}
where $l=2$. With this wave function as input and by using the
same parameters as for the $S$ wave state, we can solve the
approximate Salpeter equation, $i.e.$, only the positive energy
part solution is considered. The mass of this pure
$D$ wave state is $2672$ MeV, and the radial wave function of this
pure $D$ wave state is shown in Figure 7.

\begin{figure}
\begin{picture}(400,170)(0,-5)
\put(0,0){\includegraphics{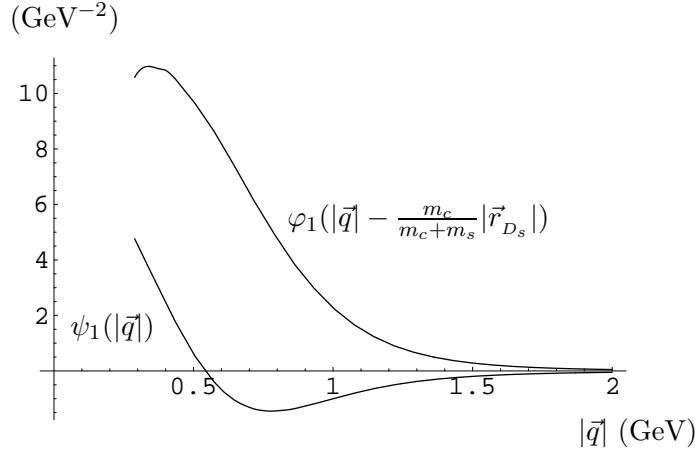}}
\put(305,0){\makebox(2,2)[l]{$|\vec q| \text{ (GeV)}$}}
\put(90,157){\makebox(2,2)[l]{(GeV$^{-2}$)}}
\put(195,80){\makebox(2,2)[l]{$\varphi_{1}(|\vec q|-\frac{m_c}{m_c
+m_s}{|\vec r_{_{D_{s}}}|})$}}
\put(113,40){\makebox(2,2)[l]{$\psi_{1}(|\vec q |)$}}
\end{picture}
\caption{The overlapping part of radial wave function
$\psi_{1}(|\vec q |)$ and $\varphi_{1}(|\vec q|-\frac{m_c}{m_c
+m_s}{|\vec r_{_{D_{s}}}|})$ when the momentum of $\varphi_{1}$
has the biggest shift $\frac{m_c}{m_c +m_s}{|\vec r_{_{D_{s}}}|}$.
}
\end{figure}

\begin{figure}
\begin{picture}(400,170)(0,-5)
\put(0,0){\includegraphics{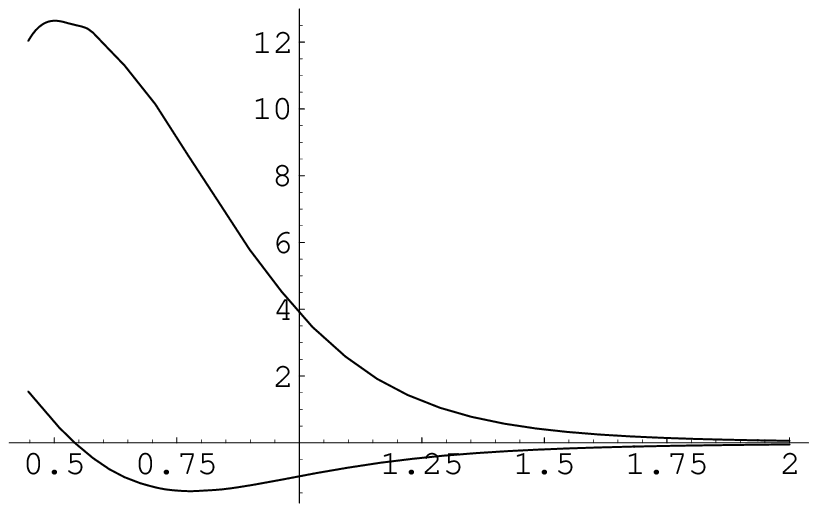}}
\put(305,0){\makebox(2,2)[l]{$|\vec q| \text{ (GeV)}$}}
\put(160,157){\makebox(2,2)[l]{(GeV$^{-2}$)}}
\put(185,67){\makebox(2,2)[l]{$\varphi_{1}(|\vec q|-\frac{m_c}{m_c
+m_u}{|\vec r_{_{D^{0}}}|})$}}
\put(107,41){\makebox(2,2)[l]{$\psi_{1}(|\vec q |)$}}
\end{picture}
\caption{The overlapping part of radial wave function
$\psi_{1}(|\vec q |)$ and $\varphi_{1}(|\vec q|-\frac{m_c}{m_c
+m_u}{|\vec r_{_{D^{0}}}|})$ when the momentum of $\varphi_{1}$
has the biggest shift $\frac{m_c}{m_c +m_u}{|\vec
r_{_{D^{0}}}|}$.}
\end{figure}

\begin{figure}
\begin{picture}(400,170)(0,-5)
\put(0,0){\includegraphics{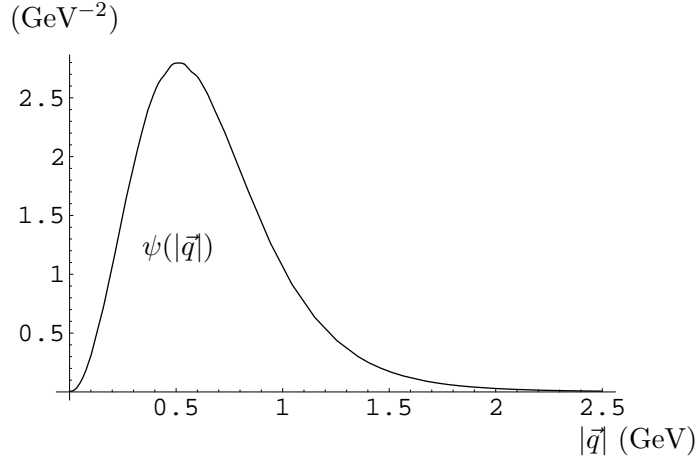}}
\put(305,-3){\makebox(2,2)[l]{$|\vec q| \text{ (GeV)}$}}
\put(90,157){\makebox(2,2)[l]{(GeV$^{-2}$)}}
\put(140,70){\makebox(2,2)[l]{$\psi(|\vec q |)$}}
\end{picture}
\caption{Radial wave function $\psi(|\vec q |)$ of
$D_{sJ}^{+}(2632)$ if it is a $D$ wave state. }
\end{figure}

To estimate the matrix element in this case, we calculate the overlapping integral of
initial and final wave functions. There is only one term, which can give a
non-zero contribution:
\begin{eqnarray}
\int \frac{d^3{\vec q}}{(2\pi)^3}\varphi_{i}({ \vec
q}-\frac{m_c}{m_c +m_q}{\vec r })\psi_{nll_z}(\vec
q)q^{\alpha}q^{\beta}f({ \vec q}^2)=\epsilon^{\alpha\beta}(l_z)c,
\end{eqnarray}
where $i=1,2$; $c$ is a constant; $f({ \vec q}^2)$ is a shorthand
notation for the terms which are functions of ${ \vec q}^2$;
$\epsilon^{\alpha\beta}(l_z)$ is the polarization tensor, which
couples with the polarization vector $\epsilon^{\gamma}(s_z)$ to
make the total polarization vector $\epsilon^{\alpha}(j_z)$:
$$
\sum_{s_z,l_z}\epsilon^{\alpha}(s_z)\epsilon^{\beta\gamma}(l_z)
<1s_{z};ll_{z}\mid 1j_{z}>=-\sqrt{\frac{3}{20}}
\Bigg[\frac{2}{3}\epsilon^{\alpha}(j_z)
(-g^{\beta\gamma}+\frac{P^{\beta}P^{\gamma}}{M^2})
$$
\begin{eqnarray}
-\epsilon^{\beta}(j_z)(-g^{\alpha\gamma}+\frac{P^{\alpha}P^{\gamma}}{M^2})-
\epsilon^{\gamma}(j_z)(-g^{\alpha\beta}+\frac{P^{\alpha}P^{\beta}}{M^2})\Bigg]\,.
\;\;\;\;\; (l=2)
\end{eqnarray}
Then, we obtain the numerical values for the decay widths when
$D_{sJ}^{+}(2632)$ is a pure $D$ wave state ($J^P=1^{-}$):
\begin{equation}
\Gamma(D^{+}_{sJ}\rightarrow D^{+}_s\eta)=3.88 \pm 0.32 \text{
MeV},
\end{equation}
\begin{equation}
\Gamma(D^{+}_{sJ}\rightarrow D^{0}K^{+})=23.3 \pm 3.2 \text{ MeV},
\end{equation}
\begin{equation}
\Gamma(D^{+}_{sJ}\rightarrow D^{+}K^{0})=21.5 \pm 3.1 \text{ MeV}.
\end{equation}
And the corresponding ratios are
\begin{equation}
\Gamma(D^{+}_{sJ}\rightarrow
D^{0}K^{+})/\Gamma(D^{+}_{sJ}\rightarrow D^{+}_{s}\eta) \simeq
\Gamma(D^{+}_{sJ}\rightarrow
D^{+}K^{0})/\Gamma(D^{+}_{sJ}\rightarrow D^{+}_{s}\eta) \simeq
6.0\pm 0.5\,,
\end{equation}
which show that the pure $D$ wave assumption does not fit the experimental
data obviously.

\begin{figure}
\begin{picture}(400,170)(0,-5)
\put(0,0){\includegraphics{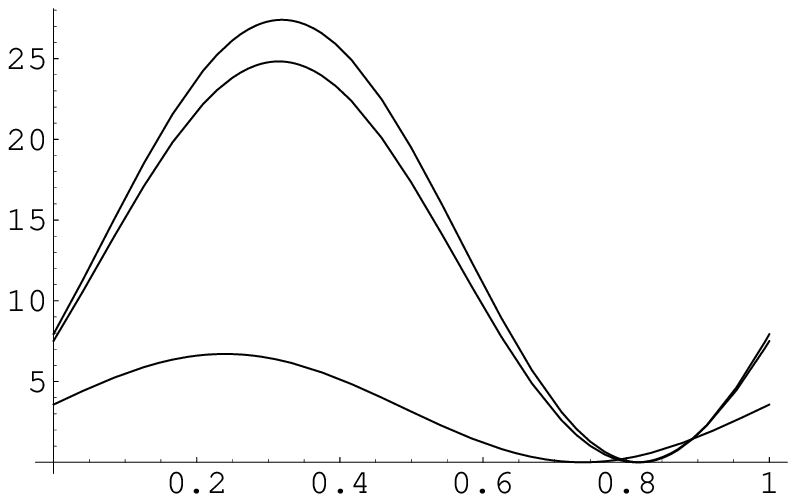}}
\put(300,-10){\makebox(2,2)[l]{$\theta (\pi)$}}
\put(90,155){\makebox(2,2)[l]{(MeV)}}
\put(132,150){\makebox(2,2)[l]{$\Gamma(D^{+}_{sJ}\rightarrow
D^{0}K^{+})$}}
\put(132,80){\makebox(2,2)[l]{$\Gamma(D^{+}_{sJ}\rightarrow
D^{+}K^{0})$}}
\put(140,55){\makebox(2,2)[l]{$\Gamma(D^{+}_{sJ}\rightarrow
D^{+}_s\eta)$}}
\end{picture}
\caption{The decay width with different mixing angle $\theta$,
when initial state mass is $2632$ MeV and it is an $S-D$ mixing
state.}\label{mixing}
\end{figure}

We now consider the possibility of the $S-D$ mixing with a mixing angle $\theta$:
\begin{eqnarray}
|D^{*}_{sJ}\rangle &=& \cos\theta|2S\rangle+\sin\theta|1D\rangle\,,\nonumber
\\
|D^{**}_{sJ}\rangle &=& -\sin\theta|2S\rangle+\cos\theta|1D\rangle\,.
\end{eqnarray}
We show the decay widths for
the channels $D^{+}_{sJ}\rightarrow D^{+}_{s}\eta,~
D^{+}_{sJ}\rightarrow D^{0}K^{+}$ and $D^{+}_{sJ}\rightarrow
D^{+}K^{0}$  as a function of the mixing angle $\theta$ in Figure \ref{mixing}.
Numerically
it is easy to find that when the mixing
angle is taken to be $\theta=0.807\pi$,  the relative
branching ratio
\begin{equation}
\Gamma(D^{+}_{sJ}\rightarrow
D^{0}K^{+})/\Gamma(D^{+}_{sJ}\rightarrow D^{+}_{s}\eta)=0.14
\end{equation}
can fit the SELEX data.
We also predict other relative branching ratios
\begin{eqnarray}
&\Gamma(D^{+}_{sJ}\rightarrow
D^{+}K^{0})/\Gamma(D^{+}_{sJ} \rightarrow D^{+}_{s}\eta) \simeq 0.05, \nonumber \\
{\rm and}~~~~~
&\Gamma(D^{+}_{sJ}\rightarrow
D^{+}K^{0})/\Gamma(D^{+}_{sJ} \rightarrow D^{0}K^+) \simeq 0.37,
\end{eqnarray}
with a rather small total decay width
\begin{equation}
\Gamma =
\Gamma(D^{+}_{sJ}\rightarrow D^{+}_{s}\eta)+
\Gamma(D^{+}_{sJ}\rightarrow D^{0}K^{+})+
\Gamma(D^{+}_{sJ}\rightarrow D^{+}K^{0})+\cdots
\simeq 0.35~~{\rm MeV},
\end{equation}
which is quite smaller than the experimental bound for the full width ($\leq 17$ MeV).
It is
interesting to note that in fact there is another solution for
the mixing angle, $\theta=0.837\pi$, which can also fit the data of
the relative branching ratio. For this solution, accordingly, the
relative branching ratios
$\Gamma(D^{+}_{sJ}\rightarrow D^{+}K^{0})/\Gamma(D^{+}_{sJ}\rightarrow D^{+}_{s}\eta)\simeq 0.20$
and
$\Gamma(D^{+}_{sJ}\rightarrow D^{+}K^{0})/\Gamma(D^{+}_{sJ}\rightarrow D^{0}K^+)\simeq 1.4$
are predicted with a slightly larger total decay width
$\Gamma\simeq 0.82$ MeV. The two solutions present a different
relative aspect on the two decay channels $D^{+}_{sJ}\rightarrow
D^{0}K^{+}$ and $D^{+}_{sJ}\rightarrow D^{+}K^{0}$.

As a final note, we only estimated the OZI allowed strong decay
widths $D^{+}_{sJ}\rightarrow D_{s}\eta$ and
$D^{+}_{sJ}\rightarrow DK$ assuming that $D^{+}_{sJ}$ is an
$1^{-}$ vector state. However, there are some other mechanisms
through which the $1^{-}$ vector state can decay to the same final
states, which we do not consider here. For example, there is the
OZI forbidden mechanism for $D^{+}_{sJ}$ decaying to $D_s \eta$
through the anomaly term in PCAC, $i.e.$ it can decay through two
gluons, $D^{+}_{sJ}\rightarrow D_{s}+gg$, $gg\rightarrow \eta$,
which is a decay similar to $\Psi^{'}\to J/\Psi\eta$ (with decay
width about $8.9$ keV). As is well known, the
OZI forbidden processes are small and cannot improve sufficiently
the relative ratio to agree with the SELEX data. Therefore, we do not
include the processes in our analysis.

As summary, according to the data from SELEX and our estimates
made here, it is not likely that the $D_{sJ}^{+}(2632)$ is the
pure $S$-wave $1^{-}$ vector state. Only the $S-D$ wave mixing
assumption may explain the relative branching ratio
$\Gamma(D^{+}_{sJ}\rightarrow
D^{0}K^{+})/\Gamma(D^{+}_{sJ}\rightarrow D^{+}_{s}\eta)=0.14$,
whereas the assumption seems to give a rather small total decay
width (to be compared with the experimental bound $\Gamma\leq 17$
MeV). Generally speaking, our results (based on PCAC and low
energy theorem) are qualitatively similar to those of
Ref.\cite{feclose} (based on $^3P_0$ model). We find that the node
structure for the first radially excited state of the $1^{-}$
ground state $D^{*}_{s}(2112)$ can have a favorable effect on the
relative ratio of the decay widths but it can not be so big as
indicated by the data. The $S-D$ wave mixing may help to explain
the data, however, more precise measurements on the relative ratio
$\frac{\Gamma(D^{+}_{sJ}\rightarrow
D^{+}K^{0})}{\Gamma(D^{+}_{sJ}\rightarrow D^{+}_{s}\eta)}$ (and/or
$\frac{\Gamma(D^{+}_{sJ}\rightarrow
D^{+}K^{0})}{\Gamma(D^{+}_{sJ}\rightarrow D^{0}K^+)}$) as well as
the total decay width are crucial. Therefore, the explanation of
$D^{+}_{sJ}(2632)$ as the first radially excited state of the
$1^{-}$ ground state $D^{*}_{s}(2112)$ is too early to conclude.
\\

\noindent
{\Large \bf Acknowledgements}

We would like to thank G. Cvetic for careful reading of the manuscript and his
valuable comments.
This work of C.H.C. and G.L.W. was supported by the
National Natural Science Foundation of China (NSFC). The work of
C.S.K. was supported in part by  CHEP-SRC Program and  in part by
Grant No. R02-2003-000-10050-0 from BRP of the KOSEF. The work of
G.L.W. was also supported in part by BK21 Program, and in part by
Grant No. F01-2004-000-10292-0 of KOSEF-NSFC International
Collaborative Research Grant.

\end{document}